\RequirePackage{fix-cm}

\documentclass[twocolumn,epjc3]{svjour3}

\RequirePackage{graphicx}
\usepackage[nottoc,numbib]{tocbibind}
\usepackage{amsmath,graphicx,color,hyperref}
\usepackage{appendix}
\newcommand{\trento}{T\raisebox{-0.5ex}{R}ENTo}

\journalname{Eur. Phys. J. A}

\begin{document}

 \title{Many-body correlations for nuclear physics across scales:\\from nuclei to quark-gluon plasmas to hadron distributions}

 \author{Giuliano Giacalone\thanksref{addr1,e1}}
 
 \thankstext{e1}{giacalone@thphys.uni-heidelberg.de}

\institute{ Institut f\"ur Theoretische Physik, Universit\"at Heidelberg,
 Philosophenweg 16, 69120 Heidelberg, Germany \label{addr1} }

\date{}

\maketitle

\begin{abstract}
It is an experimental fact that multi-particle correlations in the final states of high-energy nucleus-nucleus collisions are sensitive to collective correlations of nucleons in the wave functions of the colliding nuclei. Here, I show that this connection is more direct than it intuitively seems. With an energy deposition scheme inspired by high-energy quantum chromodynamics, and within a linearized description of initial-state fluctuations in the quark-gluon plasma, I exhibit relations between $N$-particle correlations in the final states of nuclear collisions and $N$-nucleon density distributions in the colliding nuclei. This result formally justifies the sensitivity of the outcome of high-energy collisions to features such as nuclear deformations. It paves the way, thus, to systematic studies of the impact of state-of-the-art nuclear interactions in such processes.
\end{abstract}

\tableofcontents

\section{Introduction}

Multi-particle correlations in the final states of ultrarelativistic nuclear collisions provide crucial insights about the initial condition and the dynamics of the quark-gluon plasma (QGP \cite{Braun-Munzinger:2007edi,Teaney:2009qa,Shuryak:2014zxa,Busza:2018rrf}) formed in such processes. For this reason they have been extensively studied at the BNL Relativistic Heavy Ion Collider (RHIC) and the CERN Large Hadron Collider (LHC) \cite{ATLAS:2014ndd,ALICE:2014dwt,ATLAS:2014qxy,ALICE:2016kpq,ALICE:2017fcd,CMS:2017glf,ALICE:2018rtz,PHENIX:2018lfu,ATLAS:2019pvn,ATLAS:2019peb,ALICE:2021adw,ALICE:2021klf,ALICE:2021gxt,ALICE:2022xhd,STAR:2022gki,ATLAS:2022dov,ALICE:2023lwx}. In the limit of central collisions, where the nuclei overlap nearly head-on, these measurements are strongly sensitive to collective spatial correlations of nucleons in the colliding nuclear wave functions.  In a classical treatment where correlations are encapsulated in \textit{intrinsic shapes} \cite{BMbook}, high-energy experiments have indeed provided complementary evidence of the quadrupole, octupole, and hexadecapole deformations of several species \cite{STAR:2015mki,ALICE:2018lao,CMS:2019cyz,ATLAS:2019dct,STAR:2021mii,ALICE:2021gxt,ATLAS:2022dov}. These findings support a picture of high-energy scattering as an imaging process giving access to correlated (including up to $A$-body correlations) spatial distributions of nucleons in the ground states of the colliding ions \cite{Bally:2022vgo}, and have attracted considerable attention in the theoretical community in the past couple of years (see e.g. \cite{Giacalone:2021uhj,Bozek:2021zim,Xu:2021vpn,Summerfield:2021oex,Xu:2021qjw,Giacalone:2021udy,Jia:2021wbq,Jia:2021tzt,Bally:2021qys,Jia:2021qyu,Zhang:2021kxj,Jia:2021oyt,Xu:2021uar,Nijs:2021kvn,Rong:2022qez,Liu:2022kvz,Jia:2022iji,Zhang:2022fou,Zhao:2022uhl,Jia:2022qrq,Magdy:2022cvt,Jia:2022qgl,Nie:2022gbg,Liu:2022xlm,Zhao:2022mce,Liu:2023qeq,Cheng:2023ciy,Dimri:2023wup,Bhatta:2023cqf,Samanta:2023tom,Bally:2023dxi,Mehrabpour:2023ign,Ryssens:2023fkv,Luzum:2023gwy,Mantysaari:2023jny,Bui:2023gum,Giacalone:2023cet,Serenone:2023zbn,Wang:2023yis}).

One is naturally led to ask what features of the strong nuclear force experiments at high energy enable us to probe. This is especially compelling in the context of modern \textit{ab initio} approaches to the nuclear many-body problem \cite{Hergert:2020bxy,Gandolfi:2020pbj,Soma:2020xhv,Lahde:2019npb,Ekstrom:2022yea}, where the nuclear force emerges in an effective theory of low-energy QCD, dubbed chiral effective field theory \cite{Machleidt:2016rvv,Hammer:2019poc,Epelbaum:2019kcf,Piarulli:2019cqu}.  To elucidate the complementarity of low- and high-energy experiments, it would be thus desirable to perform systematic implementations of state-of-the-art nuclear theory predictions in simulations of high-energy collisions. More concretely, it would be insightful to assess how the outcome of the simulations changes under parameter variations, different resolution scales and truncations of the chiral effective field theory expansion.

Connection between more or less advanced \textit{ab initio} calculations of nuclear structure and high-energy collisions has been made in the past to highlight the effects of nuclear geometry and nucleon-nucleon correlations in collisions of light nuclei, from deuteron to $^{16}$O \cite{Nagle:2013lja,Lim:2018huo,Rybczynski:2019adt,Summerfield:2021oex,Nijs:2021clz}. In these works, the Schrödinger equation is solved via Monte Carlo methods which give access to fully-correlated nucleon configurations sampled from the $A$-body nuclear wave function. While these results provide state-of-the-art information for the simulation of the collider processes, we have at present no understanding in regards to what properties of the sampled  wave functions are important for the phenomenology. This is also a open question in nuclear structure itself, as what precisely drives nuclear deformations in \textit{ab initio} calculations based on chiral effective field theory is yet to be fully clarified \cite{Ekstrom:2023nhc}.  Likely, the most prominent deformations are captured by  2-, 3- and possibly 4-nucleon correlations in the considered ground states. At high energy, what is missing is a theoretical description connecting multi-hadron correlation observables to $N$-nucleon correlations in the colliding ions. This would pave the way to more systematic analyses connecting nuclear structure predictions to high-energy collisions, as $N$-nucleon densities may be simpler to obtain than correlated $A$-body configurations.

In this paper, a step is taken in this direction. I show that, indeed, under certain conditions $N$-hadron correlations in the final states of nuclear collisions (whose definition I recall in Sec.~\ref{sec:2}) can be directly linked to $N$-nucleon densities in the colliding ions. This is achieved in a two-step procedure.  First, in Sec.~\ref{sec:3} I invoke a linearized description of initial-state fluctuations in the QGP to relate final-state hadron correlations to correlation functions of the fluctuating energy density field characterizing the QGP on an event-by-event basis.  In a second step, discussed in Sec.~\ref{sec:4}, a simple and yet realistic parametrization of the QGP energy density is employed, which involves the product of nuclear profiles. This model leads then to a straightforward link between energy-density correlators in the QGP and many-body densities in the colliding nuclei, connecting thus nuclear structure properties and final-state observables (including the output of photon-mediated interactions at high energy, as discussed in Sec.~\ref{sec:5}). In Sec.~\ref{sec:6}, comprehensive numerical tests are carried out to assess the validity of the approximations underlying the present analysis and their applicability. The corresponding figures are reported in Appendix~\ref{sec:appA}. Section~\ref{sec:7} concludes the paper with a summary and an outlook on possible research directions opened by this study.

\section{Multi-particle correlations in heavy-ion collisions}
\label{sec:2}

\begin{figure}[t]
    \centering
    \includegraphics[width=.85\linewidth]{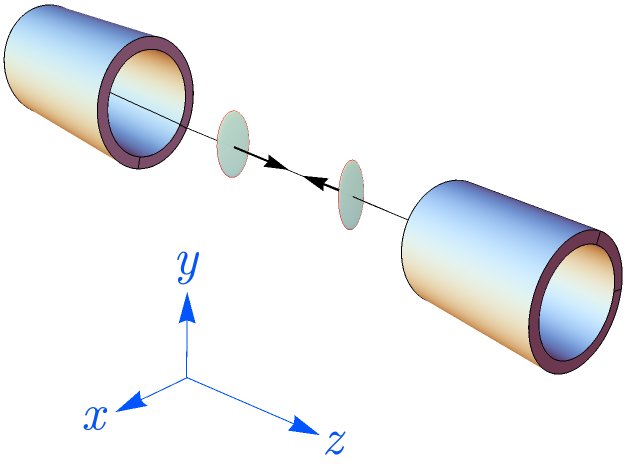}
    \caption{Sketch of an ultrarelativistic collision between nuclei in the lab frame, where the nuclei are flattened along the beam direction, $z$, by a large Lorentz factor. The coordinates $x$ and $y$ define the transverse plane. The collision occurs at zero impact parameter, with the center-of-mass of each nucleus lying at $x=y=0$.}
    \label{fig:0}
\end{figure}

The detectable outcome of a nuclear collision at high energy is a hadron spectrum differential in momentum:
\begin{equation}
    \frac{dN}{d\phi p_t dp_t d\eta},
\end{equation}
where $p_t$ is the magnitude of the momentum in the \textit{transverse plane}, orthogonal to the collision axis, $\phi$ is its azimuthal direction, while $\eta$ is the so-called \textit{pseudorapidity}, related to the longitudinal component of the momentum via its polar angle of emission relative to the beam pipe, $\theta = 2 \arctan \left ( e^{-\eta} \right )$, such that $\eta=0$ implies an emission orthogonal to the beam direction at $z=0$ in Fig.~\ref{fig:0}. In the lab frame, the part of the wave function of the colliding nuclei that determines the spatial positions of the nucleons, or, in general, of the degrees of freedom having large values of the Bjorken-$x$ variable (such as valence quarks) is squeezed in beam direction by a Lorentz factor, $\gamma$, which at top BNL RHIC and the CERN LHC energy satisfies $\gamma>100$. The longitudinal extent of the nuclei is therefore negligible and the collision is that of two flat disks (see Fig.~\ref{fig:0}). All the relevant information about the collision dynamics is carried as a consequence by the hadron spectrum measured at \textit{midrapidity}, on which I shall focus:
\begin{equation}
 \frac{dN}{d\phi p_t dp_t} =  \frac{dN}{d\phi p_t dp_t d\eta}\biggl|_{\eta=0}.
\end{equation}
The total yield of hadrons in a collision event is:
\begin{equation}
    N_{\rm ch} = \int d\phi p_t dp_t~ \frac{dN}{d\phi p_t dp_t},
\end{equation}
coming from the contribution of several species (typically 80\% pions, 15\% kaons, and 5\% heavier particles). 

At high enough energy, the rescattering of partons in the interaction region leads on a time scale of order 1 fm/$c$ to the formation of a system that is close to thermal equilibrium \cite{Schlichting:2019abc}, the QGP, a near-perfect fluid characterized by the equation of state of hot QCD \cite{Bernhard:2019bmu,Gardim:2019xjs}. One of the main goals of the high-energy nuclear collision programs is indeed the characterization of this medium from experiments \cite{ALICE:2022wpn,Arslandok:2023utm}.

The hydrodynamic expansion affects mainly the production of \textit{soft} particles sitting at low values of $p_t$, typically $p_t < 2$ GeV. Precise information about the flow of the QGP can be reconstructed from global properties of the observed spectra. One such property is the average magnitude of the hadron momenta,
\begin{equation}
\label{eq:mpt}
   [p_t] = \frac{1}{N_{\rm ch}} \sum_{i=1}^{N_{\rm ch}} p_{t,i}  
\end{equation}
quantifying the explosiveness of the QGP expansion. Second, one looks at the azimuthal distribution of the produced hadrons via a Fourier decomposition \cite{Heinz:2013th}, 
\begin{equation}
     \frac{dN}{d\phi p_t dp_t} = \frac{dN}{p_t dp_t} \sum_{n=-\infty}^{\infty} V_n(p_t) e^{i n \phi}, \hspace{10pt}|V_n|=v_n,
\end{equation}
where the complex harmonics, $V_n (p_t)$, dubbed coefficients of \textit{anisotropic flow}, encode the anisotropy of the particle emission. In their $p_t$-integrated form, they read: 
\begin{equation}
\label{eq:Vn}
    V_n = \frac{1}{N_{\rm ch}} \sum_{i=1}^{N_{\rm ch}} e^{-in\phi_i}.
\end{equation}
In spite of the abundant production of hadrons,  well-defined values of $V_n$ and $[p_t]$ on an event-by-event basis can not be obtained, due to large statistical fluctuations associated with the finite $N_{\rm ch}\sim \mathcal{O}(1000)$. To measure meaningful quantities, experiment sort the collected collisions (or \textit{events}) into classes, and evaluate averages of $V_n$ and $[p_t]$ from these larger samples. Suppose an event class contains $N_{\rm event}$ collisions producing $N_{\rm ch}$ hadrons on average. The effective number of particles used in the calculation of observables becomes of order $N_{\rm ch}\times N_{\rm event}$, which is infinite in practice.

The simplest observable is the mean value of the average momentum in the event class,
\begin{equation}
    \langle [p_t] \rangle_{\rm ev} = \frac{1}{N_{\rm event} } \sum_{{\rm ev}=1}^{N_{\rm event}}  [p_t],
\end{equation}
where I have introduced the notation
\begin{equation}
    \left \langle \ldots \right \rangle_{\rm ev} = \frac{1}{N_{\rm event}} \sum_{{\rm ev}=1}^{N_{\rm event}} \ldots~~.
\end{equation}
Fluctuations of $[p_t]$ are also important \cite{Broniowski:2009fm,Bozek:2012fw,Bozek:2017elk,Giacalone:2020lbm,Samanta:2023amp}. The variance, ${\rm var}([p_t])$, and the skewness, ${\rm skew}([p_t])$, of the distribution of $[p_t]$ in the event class can be obtained from correlations of momenta \cite{STAR:2005vxr,STAR:2013sov,ALICE:2014gvd,STAR:2019dow,Saha:2023hcy}:
\begin{align}
\label{eq:pt-var}   & {\rm var}([p_t]) = \left \langle  \frac{\sum_{i\neq j} (p_i - \langle [p_t]_{\rm ev} \rangle)(p_j - \langle [p_t]_{\rm ev} \rangle)}{N_{\rm ch,ev}(N_{\rm ch,ev}-1)}  \right \rangle_{\rm ev} , \\
\nonumber   & {\rm skew}([p_t]) = \\
\label{eq:pt-skew}   & \left \langle  \frac{\sum_{i\neq j \neq k} (p_i - \langle [p_t]_{\rm ev} \rangle)(p_j - \langle [p_t]_{\rm ev} \rangle)(p_k - \langle [p_t]_{\rm ev} \rangle)}{N_{\rm ch,ev}(N_{\rm ch,ev}-1)(N_{\rm ch,ev}-2)}  \right \rangle_{\rm ev} ,
\end{align}
where $N_{\rm ch, ev}$ is the event-to-event multiplicity. These observables represent two examples of the aforementioned multi-particle correlations constructed in the final state of high-energy nuclear collisions. Specifically, Eq.~(\ref{eq:pt-var}) is a two-particle correlation, while Eq.~(\ref{eq:pt-skew}) is a three-particle correlation. Note that the sums over particle pairs $(i,j)$ in  Eq.~(\ref{eq:pt-var}) and over all particle triplets $(i,j,k)$ in Eq.~(\ref{eq:pt-skew}) excludes all double-counting of the same particles, such that physically-uninteresting self-correlations are not included in the observable. 

Moving on to the anisotropic flow coefficients, one has to first note that an average of $V_n$ in the event class must be zero, $\langle V_n \rangle_{\rm ev}=0$, because the orientation of the anisotropy of the particle emission is random on an event-by-event basis. Hence, one can only measure the mean squared modulus of the Fourier harmonic, which cancels the random phase,
\begin{equation}
\label{eq:vn2p}
    v_n\{2\}^2 \equiv  \langle V_n V_n^* \rangle_{\rm ev}  = \left \langle  \frac{\sum_{i \neq j} e^{in(\phi_i-\phi_j)}}{ N_{\rm ch,ev}(N_{\rm ch,ev}-1)} \right \rangle_{\rm ev},
\end{equation}
corresponding to a two-particle azimuthal correlation. Higher-order moments of the $v_n^2\equiv V_nV_n^*$ distribution can be constructed by taking further azimuthal angles in the average, though I do not consider this possibility here.  In the present study I need, however, a three-particle covariance \cite{Bozek:2016yoj,ATLAS:2019pvn,ALICE:2021gxt,ATLAS:2022dov},
\begin{equation}
\label{eq:vn-pt3p}
   {\rm cov}([p_t],v_n^2) =   \left \langle \frac{\sum_{i\neq j \neq k} (p_i - \langle [p_t]_{\rm ev} \rangle)e^{in(\phi_j-\phi_k)}}{N_{\rm ch,ev}(N_{\rm ch, ev}-1)(N_{\rm ch, ev }-2)} \right \rangle_{\rm ev}.
\end{equation}
quantifying the statistical correlation between the explosiveness and the anisotropy of the particle flow \cite{Bozek:2020drh,Schenke:2020uqq,Giacalone:2020dln}.

This clarifies what multi-particle correlation measurements represent and what experimental information they involve. The goal of this manuscript is to relate these observables to multi-nucleon correlations in the wave functions of the colliding nuclei. The next step is to discuss the physical origin of $[p_t]$, $V_n$, and their fluctuations to relate early-time properties of the QGP to experimental data.

\section{Origin of flow fluctuations in heavy-ion collisions}

\label{sec:3}

Before proceeding, I stress that this study deals with multi-particle correlations that are sourced at the level of the initial conditions of the QGP. The key realization is that, even in collisions at fixed impact parameter, the QGP is shaped by a distribution of energy density whose geometry fluctuates on an event-by-event basis. The hydrodynamic expansion is driven by pressure-gradient forces. The flow velocity and its anisotropy are thus determined by the initial spatial distribution of pressure gradients. If this geometry is different in every realization of the QGP, then each expansion leads to a different flow pattern.

Additional sources of fluctuations associated with the dynamics of particlization of the QGP to detectable hadrons are present in the picture and can lead to contributions to the multi-particle observables introduced in the previous section.  These correlations go under the name of \textit{non-flow} contributions, and are routinely suppressed in the considered event samples with appropriate experimental techniques \cite{Jia:2017hbm}.

\subsection{Properties of the energy-density field}

A collision event yields a distribution of energy density, $\epsilon({\bf x})$, in the transverse plane, parametrized as ${\bf x}=(x,y)$ (see Fig.~\ref{fig:0}). Concerning the selection of event classes, experimentally this is typically done by grouping together collisions that present the same value of charged-particle multiplicity, $N_{\rm ch}$, in the final state. At ultrarelativistic energy, the energy of a particle equals its momentum, therefore, the average momentum $[p_t]$ measures the energy per particle. In view of this, in a sample of events having the same number of particles, $[p_t]$ is essentially determined by the amount of energy stuffed in the collision area \cite{Gardim:2020sma,Giacalone:2020dln,Samanta:2023amp}. This is the integral of the density field,
\begin{equation}
    E = \int_{\bf x} \epsilon({\bf x}).
\end{equation}
Similarly, the Fourier harmonics $V_n$ are sourced by the anisotropy that characterizes the spatial distribution of energy density ($|{\bf x}|\equiv \sqrt{x^2+y^2}$, $\phi_x={\rm atan2}(y/x)$):
\begin{equation}
\label{eq:epsn}
    \mathcal{E}_n = - \frac{\int_{\bf x} \epsilon({\bf x})~|{\bf x}|^n e^{in\phi_x} }{\int_{\bf x} 
 \epsilon({\bf x}) ~|{\bf x}|^n} ,
\end{equation}
in the sense that if $\mathcal{E}_n=0$, then the hydrodynamic expansion leads to $V_n=0$. Note that for $n=2$ and $n=3$ (on which I focus here), the multipole moments in the numerator of Eq.~(\ref{eq:epsn}) can be rigorously derived from a cumulant expansion of $\epsilon({\bf x})$, and shown to represent the relevant measures of $n$th order anisotropy associated with long wavelength modes of the system \cite{Teaney:2010vd}.

\begin{figure*}
    \centering
    \includegraphics[width=.9\linewidth]{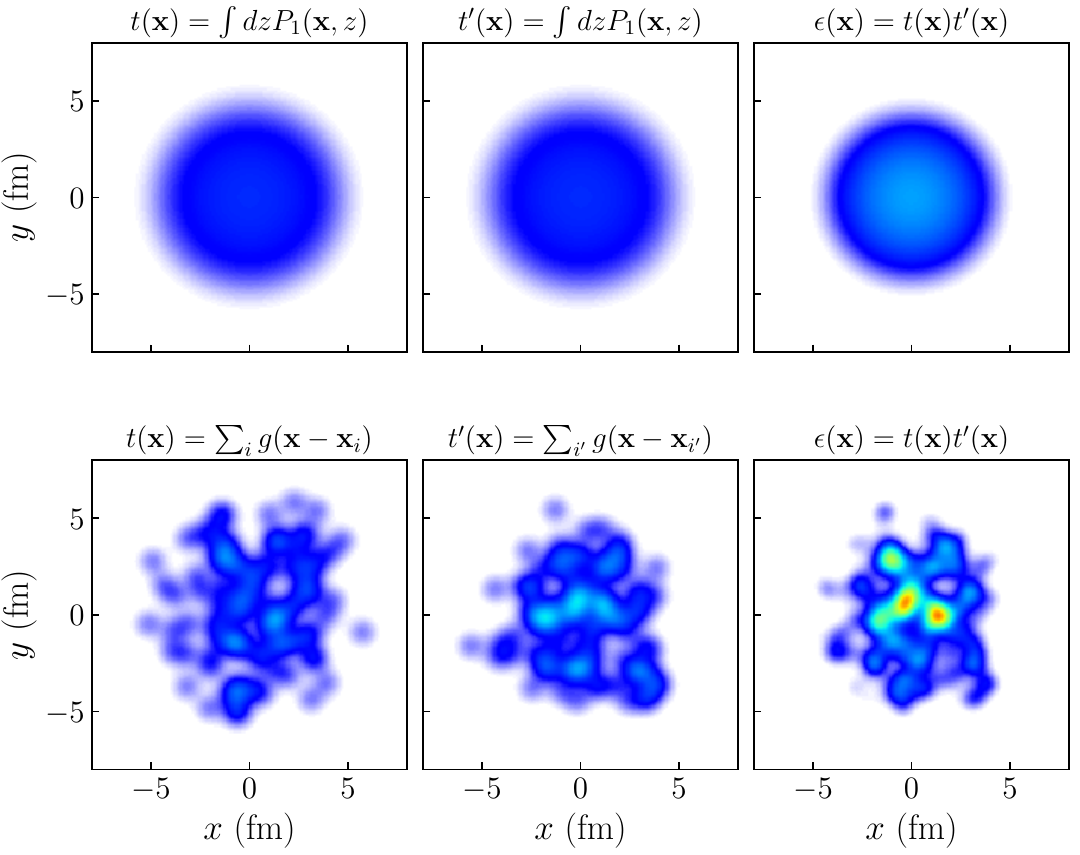}
    \caption{Energy density (in arbitrary units) deposited in the transverse plane in the collision of two nuclei with mass number $A=96$ at zero impact parameter, $b=0$. The energy density, $\epsilon({\bf x})$, is given by the product of the transverse profiles of the two colliding nuclei, $t({\bf x})$ and $t^\prime({\bf x})$, respectively, at the time of scattering. \\
     \textbf{Top:} the colliding nuclei are identified with spin- and isospin-integrated one-nucleon densities, $P_1({\bf x},z)$, integrated over $z$ to include the effect of a Lorentz contraction. Here a Woods-Saxon profile is used for $P_1({\bf x},z)$, as in Eq.~(\ref{eq:WSfig1}). The resulting energy density profile (rightmost panel) is consequently a smooth and isotropic function over the plane.    \\ \textbf{Bottom:} quantum fluctuations associated with the finite number of nucleons are introduced in the picture. The transverse nuclear profile, $t({\bf x})$, is now an individual realization of the one-body density, and is computed as the sum of $A$ Gaussian peaks, $g({\bf x}_i)$, with a size of 0.5 fm, whose center positions are distributed according to $P_1({\bf x},z)$. The product of the two transverse profiles leads to an energy density with peaks and valleys. Spatial isotropy in the plane is broken to all orders, $\mathcal{E}_n\neq 0$. The total energy of the system, $E=\int \epsilon({\bf x})$, fluctuates as a consequence on an event-by-event basis.}
    \label{fig:1}
\end{figure*}

Therefore, in the hydrodynamic paradigm, understanding the fluctuations of $[p_t]$ and $V_n$ requires knowledge of the fluctuations of the initial total energy, $E$, and of the initial spatial anisotropies, $\mathcal{E}_n$, of the QGP. The following relations are almost exact in a class of collisions at the same multiplicity,
\begin{align}
\label{eq:linresp}
  \nonumber  [p_t] &\propto E, \\
    V_n &\propto \mathcal{E}_n.
\end{align}
Consequently, similar relations can be written for the moments of the final-state quantities, 
\begin{align}
\label{eq:corres} & {\rm var}([p_t]) \propto {\rm var}(E), \\
& {\rm skew}([p_t]) \propto {\rm skew}(E), \\
& v_2\{2\}^2  \propto \varepsilon_2\{2\}^2, \\
\label{eq:corresss} & {\rm cov}([p_t],v_n^2) \propto {\rm cov} (E,\varepsilon_n^2).
\end{align}
In this way, one is able to connect the measured multi-particle correlations, from Eq.~(\ref{eq:pt-var}), (\ref{eq:pt-skew}), (\ref{eq:vn2p}), and (\ref{eq:vn-pt3p}) to statistical correlations of the quantities $E$ and $\mathcal{E}_n$ which are determined by the event-by-event fluctuations of the initial energy density field.

A concrete example makes this discussion more transparent. I construct an energy density profile, $\epsilon({\bf x})$, in two realistic models of the collision event, that also help set the notation for the later parts of this manuscript. Consider a symmetric collision of nuclei at zero impact parameter. I consider here nuclei containing $A=96$ nucleons distributed independently according to a one-nucleon density (integrated over spin and isospin), $P_1({\bf x}, z)$, to be precisely defined in Eq.~(\ref{eq:P1z}), given by a Woods-Saxon profile,   
\begin{equation}
\label{eq:WSfig1}
    P_1({\bf x}, z) \propto \frac{1}{1 + \exp \left (  \frac{r - R}{a} \right )}~,\hspace{20pt} r=\sqrt{{\bf x}^2 + z^2}~,
\end{equation}
where $R=5$ fm is the half-width radius, and $a=0.5$ fm is the surface diffuseness. 

In a first approach, I consider a collision between two  nuclei whose spatial profile is described by a smooth function in the plane given by the Lorentz-contracted one-body density, $t({\bf x})=\int dz P_1({\bf x},z)$. The upper panels of Fig.~\ref{fig:1} shows such a situation. I consider, then, that the energy density is given by the product of two such transverse nuclear profiles, $\epsilon({\bf x}) = t({\bf x})t^\prime ({\bf x})$. The resulting energy density (upper-right panel) is a smooth and isotropic function. Therefore, in a sample of such collisions one has a constant value of total energy, $E$, while spatial anisotropies vanish by construction, $\mathcal{E}_n=0$ by construction. As nothing fluctuates, all multi-particle correlations in the final state are zero following the hydrodynamic expansion. 

In a second implementation, I consider that each colliding nucleus is obtained from an individual realization of the one-body density of the system. One samples randomly and independently from $P_1({\bf x}, z)$ the coordinates of $A$ nucleons in 3D, for both nuclei. The transverse nuclear profile is then obtained from a superposition of nucleons:
\begin{equation}
\label{eq:tx}
    t({\bf x}) = \sum_{i=1}^A g({\bf x}-{\bf x}_i),
\end{equation}
where $g({\bf x})$ is a two-dimensional nucleon form factor appropriate for high-energy scattering mediated by gluons, while ${\bf x}_i$ is the nucleon center within the scattering nucleus. Note that, as one sums over all nucleons irrespective of their $z$ coordinate, the relevant density in the transverse plane is again $\int dz P_1({\bf x},z)$.  The standard choice for the high-energy gluonic form factor is a two-dimensional Gaussian
\begin{equation}
\label{eq:Gaunucl}
    g({\bf x}-{\bf x}_i) = \frac{1}{2\pi w^2} \exp \left ( - \frac{({\bf x}-{\bf x}_i)^2}{2w^2} \right ),
\end{equation}
with a nucleon size $w=0.5$ fm. In the bottom panels of Fig.~\ref{fig:1}, the two transverse nuclear profiles $t({\bf x})$ and $t^\prime({\bf x})$ are now different. As a consequence, the energy density defined via their product becomes a fluctuating field, which breaks isotropy to all orders, $\mathcal{E}_n\neq0$, such that the hydrodynamic expansions will yield anisotropic flow, $V_n$. In a sample of events of this type, then, the total energy, $E$, becomes a fluctuating quantity, and all correlations such as ${\rm var}(E)$, ${\rm skew}(E)$, $\varepsilon_n\{2\}^2$, and ${\rm cov}(E,\varepsilon_n^2)$ are nonzero.

Twenty years of phenomenological studies have established the picture provided in the bottom panels of Fig.~\ref{fig:1} as the only viable description of heavy-ion collisions. In other words, fluctuations and correlations associated with the finite number of nucleons in the colliding nuclei are essential to explain the measured multi-hadron correlations \cite{PHOBOS:2006dbo,Alver:2010gr}.  While the calculation above employs an independent-nucleon picture for the sampling of their coordinates, a real collision corresponds instead to a sampling from a correlated wave function that contains up to $A$-body correlations. Most of nuclei are indeed characterized by strong spatial correlations at the heart of  phenomena such as nuclear deformations and clustering. From the discussion of Fig.~\ref{fig:1}, one can evince that the fluctuations of the field $\epsilon({\bf x})$ are sensitive to the details of the spatial distributions of nucleons. Relating information about many-body correlations in the colliding ions to the measured multi-particle correlation observables is the primary goal of this article.

\subsection{Formalism of $N$-point correlation functions}

The next step is to relate features of the initial conditions, such as the fluctuations of $E$ and $\mathcal{E}_n$, to more fundamental properties of the energy density field. To do so, I perform a background-fluctuation splitting \cite{Blaizot:2014nia,Floerchinger:2014fta}: 
\begin{equation}
\label{eq:back_fluct}
\epsilon({\bf x}) = \bar \epsilon ({\bf x}) + \delta \epsilon ({\bf x}),    
\end{equation}
where $\bar \epsilon ({\bf x})$ is the local average of the energy density in the event sample (here events at zero impact parameter), whose integral gives the average system's energy:
\begin{equation}
    \langle E \rangle_{\rm ev} = \int_{\bf x} C_1({\bf x}), \hspace{20pt} C_1({\bf x}) \equiv \bar \epsilon({\bf x}),
\end{equation}
while $\delta \epsilon ({\rm x})$ is the fluctuation, satisfying $\langle \delta \epsilon ({\bf x}) \rangle_{\rm ev} =0$. 

I evaluate now the correlation in Eqs.~(\ref{eq:corres})-(\ref{eq:corresss}), by inserting Eq.~(\ref{eq:back_fluct}) into the expressions of the observables and then truncating at the first nontrivial order in the perturbation, $\delta \epsilon({\bf x})$. For observables related to the fluctuations of $E$, one finds the following exact expressions. The variance reads:
\begin{align}
\nonumber    &{\rm var}(E) = \int_{{\bf x},{\bf y}} C_2({\bf x},{\bf y}),
\end{align}
where I have introduced the connected 2-point function of the density field,
\begin{equation}
    C_2({\bf x},{\bf y}) \equiv \langle \delta \epsilon({\bf x}) \delta \epsilon({\bf y}) \rangle_{\rm ev} = \langle \epsilon({\bf x}) \epsilon({\bf y})\rangle_{\rm ev} - \langle \epsilon({\bf x}) \rangle_{\rm ev} \langle \epsilon({\bf y}) \rangle_{\rm ev}.
\end{equation}
Analogously, the skewness of the total energy reads:
\begin{equation}
    {\rm skew}(E) = \int_{{\bf x},{\bf y},{\bf z}} C_3({\bf x},{\bf y},{\bf z}),
\end{equation}
which involves the connected 3-point function of the density field,
\begin{equation}
    C_3({\bf x},{\bf y},{\bf z}) = \langle \delta e({\bf x}) \delta e({\bf y}) \delta e({\bf z}) \rangle_{\rm ev}.
\end{equation}
For observables involving the spatial anisotropy, I insert Eq.~(\ref{eq:back_fluct}) into Eq.~(\ref{eq:epsn}), and then expand the denominator. As I consider only collisions at zero impact parameter, the expressions are simplified by the fact that the density background is isotropic, 
\begin{equation}
    0 = \int_{{\bf x}} C_1({\bf x}) |{\bf x}|^n e^{in\phi}.
\end{equation}
The leading expression of the mean squared anisotropy involves only the two-point function of the density \cite{Blaizot:2014nia}:
\begin{align}
\label{eq:epsnC2}
 \varepsilon_n\{2\}^2 \equiv \langle \mathcal{E}_n \mathcal{E}_n^* \rangle_{\rm ev} = \frac{ \int_{{\bf x},{\bf y}} |{\bf x}|^n|{\bf y}|^n e^{in(\phi_x-\phi_y)} C_2({\bf x},{\bf y})  }{\left ( \int_{\bf x} C_1({\bf x}) |{\bf x}|^n \right )^2}.
\end{align}
Similarly, the energy-anisotropy correlator involves the connected three-point function:
\begin{align}
\label{eq:covC3}
    {\rm cov}(E,\varepsilon_n^2)   = \frac{\int_{{\bf x},{\bf y},{\bf z}} |{\bf x}|^n|{\bf y}|^n e^{in(\phi_x-\phi_y)} C_3({\bf x},{\bf y},{\bf z}) }{\left ( \int_{\bf x} C_1({\bf x}) |{\bf x}|^n \right )^2}.
\end{align}
The validity of the approximate expressions (\ref{eq:epsnC2}) and ({\ref{eq:covC3}}) will be checked through Monte Carlo simulations in Sec.~\ref{sec:6} for different collision systems.

\subsection{Discussion}

In summary, high-energy nuclear collision experiments measure event-by-event hadron distributions from which precise information about the statistical properties of $p_t$ and $V_n$, and their correlations can be quantified via  multi-particle correlation observables. In the hydrodynamic framework, these observables probe properties of the initial condition of the QGP, such as the total energy, $E$, or of the spatial anisotropies, $\mathcal{E}_n$, which fluctuate on an event-by-event basis due to quantum fluctuations in the colliding nuclei. Fluctuations and correlations of $E$ and $\mathcal{E}_n$ can in turn be related to the correlation functions $\langle \epsilon({\bf x}) \rangle_{\rm ev}$, $\langle \epsilon({\bf x}) \epsilon({\bf y}) \rangle_{\rm ev}$, etc., of the energy density field, $\epsilon({\bf x})$, from which they are computed.

\section{Correlations from the QGP to the colliding nuclei}

\label{sec:4}

I exhibit now a link between the energy-density correlation functions, $C_1({\bf x})$, $C_2({\bf x},{\bf y})$, $C_3({\bf x, \bf y, \bf z})$ and $N$-nucleon densities in the colliding nuclei. To do so, one first needs a parametrization of the density field, $\epsilon({\bf x})$.

\subsection{Glasma-inspired model of high-energy scattering}

The idea is to take an energy deposition in the transverse plane motivated by the color glass condensate effective field theory of high-energy QCD \cite{Gelis:2010nm}. Consider two nucleons described as color glass condensates colliding at very high energy. Immediately after the collision, at proper time $\tau=0^+$, the produced system, dubbed the \textit{glasma} \cite{Lappi:2006fp}, is amenable to a classical description with an expectation value of the energy density that has a simple binary-collision scaling \cite{Lappi:2006hq}: \footnote{Note that this result is nowadays textbook material, see Exercise 11.9 in Ref.~\cite{Gelis:2019yfm}.}
\begin{equation}
\label{eq:glasm}
   \left \langle  \epsilon ({\bf x}, \tau=0^+) \right  \rangle   \propto g ({\bf x}) g^\prime ({\bf x}),
\end{equation}
where the prefactors are in principle divergent at $\tau=0$ (though logarithmically, such that the energy density per unit rapidity, $\tau \epsilon({\bf x})$, is finite at $\tau=0^+$). Here $\epsilon({\bf x})$ is a component of the glasma stress-energy tensor, while $g({\bf x})$ and $g^\prime({\bf x})$ encode respectively the spatial dependence of the average density of gluons at small $x$ within the colliding nucleons, e.g., a Gaussian profile as in Eq.~(\ref{eq:Gaunucl}). In the language of the color glass condensate theory, I thus consider that the gluon density in a nucleon is proportional to its saturation scale ($Q_s$), typically obtained through the IP-Sat model \cite{Kowalski:2003hm}.

The generalization to the case of a collision of nuclei, as included in the IP-Glasma framework \cite{Schenke:2020mbo}, takes a superposition of nucleons for the overall nuclear density, akin to Eq.~(\ref{eq:tx}): \footnote{The superposition usually involves a random normalization for the nucleon profiles (so-called $Q_s$ fluctuation \cite{Schenke:2020mbo}), i.e., $t({\bf x}) = \sum_{i=1}^{A} \lambda_i g({\bf x}-\xi_i)$, where $\lambda_i$ is a random number drawn from a more or less physically-motivated distribution. This feature could be easily added as well to the present analysis.}
\begin{equation}
    t({\bf x}) = \sum_{i=1}^{A} g({\bf x}-\xi_i).
\end{equation}
where from now on I denote by $\xi_i$ the transverse coordinate of nucleon $i$ within the nucleus. The saturation scale obtained through the IP-Sat model remains to a good approximation proportional such a superposition. To achieve the scaling of Eq.~(\ref{eq:glasm}) where $g({\bf x})$ is replaced by the average nuclear profile, one can take the energy density in an event equal to a product, as done in the lower panels of Fig.~\ref{fig:1}, 
\begin{equation}
\label{eq:jazma}
    \epsilon({\bf x}, \tau=0^+) = t ({\bf x}) t^\prime ({\bf x}),
\end{equation}
where dimensionful constants are absorbed in the functions $t({\bf x})$ and $t^\prime ({\bf x})$. This is a modified binary collision scaling where the amount of deposited energy depends on the degree of overlap of the colliding nucleons. This prescription is also called IP-Jazma model \cite{Nagle:2018ybc}, and I shall follow it in this manuscript. In terms of global collision geometry properties at $\tau=0^+$, it provides a good approximation of the IP-Glasma implementation \cite{Snyder:2020rdy}. Equation~(\ref{eq:jazma}) defines in a sense the simplest, realistic parametrization of high-energy nuclear collisions.

\subsection{Energy density correlators}

Straightforward computations lead to the correlation functions of the energy density field in this model. For the local average, $C_1({\bf x})$, one has: \footnote{I consider symmetric processes where identical nuclear species are collided. It is straightforward to generalize Eq.~(\ref{eq:1p}), and all the subsequent formulas, to asymmetric collisions of nuclei with different mass numbers, $A\neq A'$.}
\begin{equation}
\label{eq:1p}
  \langle \epsilon({\bf x}) \rangle_{\rm ev} =  C_1({\bf x}) = \left \langle \sum_{i = 1}^A \sum_{ i^\prime = 1}^A g({\bf x}-\xi_i)g({\bf x}-\xi_{i^\prime}) \right \rangle_{\rm ev}.
\end{equation}
Since $i$ and $i^\prime$ label coordinates from two different nuclei, $\xi_i$ and $\xi_{i^\prime}$ are independent variables, such that
\begin{equation}
\label{eq:C1jaz}
    C_1({\bf x}) = \left \langle  \sum_{i = 1}^A  g({\bf x}-\xi_i) \right \rangle_{\rm ev}^2 = A^2 \left ( \int_{\xi_i} P_{1\perp}(\xi_i) g({\bf x}-\xi_i) \right )^2,
\end{equation}
where $P_{1\perp}(\xi_i)$ is the probability density of finding a nucleon at transverse position $\xi_i$, irrespective of the positions of the other nucleons. The precise definition of $P_{1\perp}(\xi_i)$  will be discussed below.

I evaluate now the connected two-point function of the field. Recalling  that $i$ and $i'$ label different nuclei, the average:
\begin{align}
\label{eq:2p}
\nonumber &     \langle \varepsilon({\bf x})\varepsilon({\bf y}) \rangle_{\rm ev} = \\ 
\nonumber & \left \langle \sum_{i,j=1}^{A} \sum_{i^\prime,j^\prime=1}^{A} g({\bf x}-\xi_i)g({\bf y}-\xi_j)~g({\bf x}-\xi_{i^\prime})g({\bf y}-\xi_{j^\prime})  \right \rangle_{\rm ev} =\\
& \left \langle \sum_{i,j=1}^{A} g({\bf x}-\xi_i) g({\bf y}-\xi_j) \right \rangle^2_{\rm ev},
\end{align}
involves a two-nucleon density in the transverse plane:
\begin{align}
\label{eq:C2jaz}
  \nonumber     \langle \epsilon(&{\bf x})\epsilon({\bf y}) \rangle_{\rm ev} = \biggl ( A \int_{\xi_i} P_{1\perp} (\xi_i)  g({\bf x}-{\bf \xi}_i) g({\bf y}-{\bf \xi}_i) \\ 
      & + (A^2-A) \int_{\xi_i\neq\xi_j} P_{2\perp} (\xi_i,\xi_j)  g({\bf x}-{\bf \xi}_i) g({\bf y}-{\bf \xi}_j) \biggr)^2,
\end{align}
where we separate $A$ diagonal and $A(A-1)$ off-diagonal terms \cite{Blaizot:2014wba,Gronqvist:2016hym}. From this,  the connected two-point function is obtained:
\begin{align}
 C_2({\bf x}, {\bf y}) =  \langle \varepsilon({\bf x})\varepsilon({\bf y}) \rangle_{\rm ev}  - C_1({\bf x}) C_1({\bf y}) .
\end{align}

Analogously, the evaluation of the three-point function involves a three-point correlator:
\begin{align}
\label{eq:3p}
\nonumber     &\langle \epsilon({\bf x})\epsilon({\bf y})\epsilon({\bf z}) \rangle_{\rm ev} = \\ 
\nonumber & \biggl \langle \sum_{i,j,k=1}^{A} \sum_{i',j',k'=1}^{A} g({\bf x}-{\bf \xi}_i)g({\bf y}-{\bf \xi}_{j}) g({\bf z}-{\bf \xi}_k)~ \times \\ 
& \hspace{80pt} g({\bf x}-{\bf \xi}_{i^\prime})g({\bf y}-{\bf \xi}_{j^\prime})g({\bf z}-{\bf \xi}_{k^\prime})  \biggr \rangle_{\rm ev} .
\end{align}
Again, this can be factorized as a product of two nuclei, 
\begin{align}
\nonumber  \langle \epsilon({\bf x})\epsilon({\bf y})\epsilon({\bf z}) &\rangle_{\rm ev} = \\
  &\left \langle \sum_{i,j,k=1}^{A} g({\bf x}-{\bf \xi}_i) g({\bf x}-{\bf \xi}_j) g({\bf x}-{\bf \xi}_k) \right \rangle_{\rm ev}^2,
\end{align}
which involves a transverse three-nucleon density,
\begin{align}
\nonumber     & \langle \epsilon({\bf x})\epsilon({\bf y})\epsilon({\bf z}) \rangle_{\rm ev} = \\
\nonumber     & \biggl ( A \int_{\xi_i} P_{1\perp} (\xi_i)  g({\bf x}-{\bf \xi}_i) g({\bf y}-{\bf \xi}_i) g({\bf z}-{\bf \xi}_i) \\
\nonumber      & + A(A-1) \int_{\xi_i\neq\xi_j} P_{2\perp} (\xi_i,\xi_j)  g({\bf x}-{\bf \xi}_i) g({\bf y}-{\bf \xi}_i) g({\bf z}-{\bf \xi}_j) \\
\nonumber      & + A(A-1) \int_{\xi_i\neq\xi_j} P_{2\perp} (\xi_i,\xi_j)  g({\bf x}-{\bf \xi}_i) g({\bf y}-{\bf \xi}_j) g({\bf z}-{\bf \xi}_i) \\
\nonumber      & + A(A-1)  \int_{\xi_i\neq\xi_j} P_{2\perp} (\xi_i,\xi_j)  g({\bf x}-{\bf \xi}_j) g({\bf y}-{\bf \xi}_i) g({\bf z}-{\bf \xi}_i) \\
\nonumber      & + (A^3-3A(A-1)-A)  \\
      & \hspace{30pt} \int_{\xi_i\neq\xi_j\neq\xi_k}  P_{3\perp} (\xi_i,\xi_j,\xi_k)  g({\bf x}-{\bf \xi}_i) g({\bf y}-{\bf \xi}_j) g({\bf z}-{\bf \xi}_k) \biggr)^2.
\end{align}
The connected three-point function reads then:
\begin{align}
\nonumber    & C_3({\bf x},{\bf y},{\bf z})  =    \langle \delta \epsilon({\bf x}) \delta \epsilon({\bf y}) \delta \epsilon({\bf z})  \rangle_{\rm ev} = \\
\nonumber     &  \langle \epsilon({\bf x})\epsilon({\bf y})\epsilon({\bf z}) \rangle_{\rm ev}  \\
\nonumber     & - C_1({\bf z})C_2({\bf x,y}) - C_1({\bf y})C_2({\bf x},{\bf z}) - C_1({\bf x})C_2({\bf y},{\bf z})  \\
& - C_1({\bf x})C_1({\bf y})C_1({\bf z}) . 
    \end{align}
 
\subsection{Connection to nuclear structure}

The input from nuclear structure are the transverse nucleon densities $P_{n\perp}$, for $n=1$, 2, 3. They have an elementary derivation.

The nuclear ground state is characterized by the many-body wave function:
\begin{equation}
    \Psi(\xi_1, z_1, \xi_2, z_2, \ldots, \xi_A, z_A, s_{1}, \ldots, s_A, t_1, \ldots, t_A),
\end{equation}
where $\xi_i$ is a coordinate in the $(x,y)$ plane, the Cartesian frame $(x,y,z)$ has its origin at the center of the nucleus, and $s$ and $t$ are, respectively, projections of spin and isospin. I consider an even-even nucleus with a spherically-symmetric ground state ($J=0$). The wave function satisfies the probability condition:
\begin{equation}
   1 = \sum_{s,t} \int d^2\xi_1 dz_1 \ldots d^2\xi_A dz_A \Psi \Psi^*.
\end{equation}
I am interested in marginalized $A$-nucleon densities. The one-body density is given by (where the subscript ``1'' refers to any nucleon in the system)
\begin{equation}
\label{eq:P1z}
    P_1(\xi_1,z_1) = \sum_{s,t} \int d^2\xi_2 dz_2 \ldots d^2\xi_A dz_A \Psi \Psi^*.
\end{equation}
Now, as anticipated in the calculation of Fig.~\ref{fig:1}, in high-energy scattering the $z$ component is integrated out, defining a transverse density of nucleons:
\begin{equation}
    P_{1\perp}(\xi_1) = \int dz_1 P_1(\xi_1,z_1).
\end{equation}
Analogously, the two-body and three-body densities read:
\begin{align}
\label{eq:P2z}   &  P_2(\xi_1,z_1,\xi_2,z_2) = \sum_{s,t} \int d^2\xi_3 dz_3 \ldots d^2\xi_A dz_A \Psi \Psi^*, \\
\label{eq:P3z}   &  P_3(\xi_1,z_1,\xi_2,z_2,\xi_3,z_3) = \sum_{s,t} \int d^2\xi_4 dz_4 \ldots d^2\xi_A dz_A \Psi \Psi^*,
\end{align}
along with their transverse projections:
\begin{align}
\label{eq:P2t}   & P_{2\perp}(\xi_1,\xi_2) = \int dz_1dz_2 P_2(\xi_1,z_1,\xi_2,z_2), \\
\label{eq:P3t}   & P_{3\perp}(\xi_1,\xi_2,\xi_3) = \int dz_1dz_2dz_3 P_3(\xi_1,z_1,\xi_2,z_2,\xi_3,z_3).
\end{align}
Analogous expressions give the $A$-body densities.

\subsection{Discussion}

I recap the results of the formal discussion. With an energy deposition Ansatz following Eq.~(\ref{eq:jazma}), the 1-point function of the energy density field, $C_1({\bf x})$, is only determined by the 1-nucleon transverse density, $P_{1\perp}(\xi_1)$, in the colliding nuclei. The 2-point function of the field, $C_2({\bf x},{\bf y})$ requires in addition the 2-nucleon density  distribution, $P_{2\perp}(\xi_1,\xi_2)$, while $C_3({\bf x}, {\bf y}, {\bf z})$ requires as well $P_{3\perp}(\xi_1,\xi_2,\xi_3)$. Coupled to a linearized description of energy density fluctuations and the linear hydrodynamic response of Eq.~(\ref{eq:linresp}), one arrives thus at a direct relation between experimentally observable $N$-particle correlations and the transverse nucleon densities $P_{N\perp}$. This points to a straightforward connection between low-energy nuclear structure to the outcome of high-energy experiments, and represents my main result. 

This finding motivates the following conjecture: \textit{In collisions at fixed impact parameter, $N$-point correlation functions of the energy density field at $\tau=0^+$ are solely determined by up to $N$-nucleon density distributions in the colliding nuclei}. While there does not seem to be any fundamental argument to support such a statement,  the conjecture appears to be fulfilled in the IP-Glasma model of initial conditions. There, at $\tau=0^+$ the scaling of the energy density field in the transverse plane follows Eq.~(\ref{eq:jazma}) very closely \cite{Snyder:2020rdy,Nijs:2023yab}. An additional potential source of correlations comes from the sampling of so-called color charge fluctuations \cite{Albacete:2018bbv} in the transverse plane. However, in spite of recent claims \cite{Giacalone:2019kgg,Gelis:2019vzt}, these fluctuations seem to contribute to $C_2({\bf x},{\bf y})$ only with a delta-like signal which is negligible both in correlation length and in amplitude \cite{Snyder:2020rdy}. Therefore, at $\tau=0^+$ correlations are only sourced by nucleon positions within the colliding ions. \footnote{In
the current IP-Glasma setup \cite{Schenke:2020mbo}, this is however only true across length scales larger than the nucleon size,  $w\approx0.4$ fm. At shorter scales, the inner structure of the nucleons, parametrized in the model via the inclusion of fluctuating \textit{hot spots} that source small-$x$ gluon (akin to valence quarks),  will generate further correlations in the transverse plane after the collision takes place. It will be interesting to generalize the present study to include features related to the structure of nucleons.}

\section{Digressions}

\label{sec:5}

\subsection{Properties of the energy deposition formula}

The main result of this analysis stems from the fact that the products of source profiles in Eqs.~(\ref{eq:1p}), (\ref{eq:2p}), and (\ref{eq:3p}) can be factorized in the sum of pairwise products of sources. State-of-the-art calculations of heavy-ion collisions do not implement the Jazma-type scaling of Eq.~(\ref{eq:jazma}), but rather parametrize the energy density per unit rapidity at the initial time  in a way that can be fine-tuned from experimental data. The most generic parametrization proposed in the literature is an extended version of the \trento{} Ansatz  \cite{Moreland:2014oya} for the energy density per unit rapidity at $\tau=0^+$ \cite{Nijs:2023yab}:
\begin{equation}
  \lim_{\tau\to0^+}  \tau \epsilon({\bf x},\tau) = \left ( \frac{ t({\bf x})^p +  t^\prime({\bf x})^p 
 } {2}  \right )^{q/p},
\end{equation}
where dimensionful factors have been absorbed into $t({\bf x})$ and $t^\prime ({\bf x})$. Of all combinations of $p$ and $q$, only $p=0$ leads to an energy deposition that involves the product of the transverse nuclear densities, as it can be seen from a Maclaurin expansion of the previous equation:
\begin{equation}
\label{eq:traj}
     \lim_{\tau\to0^+}  \tau \epsilon({\bf x},\tau) = \left ( t({\bf x}) t^\prime ({\bf x}) \right ) ^ {q/2 } + \mathcal{O}(p).
\end{equation} 
Remarkably, all global Bayesian analyses of heavy-ion collision data show a strong preference for $p\approx0$ \cite{Bernhard:2016tnd,Moreland:2018gsh,Bernhard:2019bmu,Nijs:2020ors,JETSCAPE:2020shq,Parkkila:2021yha,Nijs:2021clz,Nijs:2023yab}, supporting an energy density that emerges as a simple correction to the modified binary collision picture. 
The value $q \approx 4/3$ is currently favored by CERN LHC data \cite{Nijs:2023yab}. \footnote{This is so when $t({\bf x})$ and $t^\prime({\bf x})$ are determined by the \textit{participant} nucleons, i.e., nucleons that undergo at least one nucleon-nucleon interaction. I do not discuss the implications of the participant selection. I emphasize that in the IP-Glasma model participants are not selected.}

Starting from Eq.~(\ref{eq:traj}), and with the usual assumption that $t({\bf x})$ is a superposition of nucleons, at fixed impact parameter the average energy density reads (for some real coefficient $\kappa$):
\begin{equation}
    C_1 ({\bf x}) = \left \langle  \left [ \sum_{i}^A \sum_{i^\prime}^A  g({\bf x}-{\bf x}_i)g({\bf x}-{\bf x}_{i^\prime}) \right]^\kappa \right \rangle_{\rm ev}.
\end{equation}
The correlator no longer factorizes inside the sum. This implies that $C_1({\bf x})$ is determined by all nucleon densities in the colliding wavefunctions, up to $P_{A\perp}(\xi_1. \ldots, \xi_A)$.  To avoid this, a simple possibility is to explore a modified Ansatz where the power is taken only at the level of the individual nucleon products:
    \begin{equation}
    \label{eq:modA}
    C_1 ({\bf x}) = \left \langle   \sum_{i}^A \sum_{i^\prime}^A  \left [ g({\bf x}-{\bf x}_i)g({\bf x}-{\bf x}_{i^\prime}) \right]^{\kappa^\prime} \right \rangle_{\rm ev}.
\end{equation}
For Gaussian nucleons, this is a rescaling of the nucleon width parameter, $w$. This prescription enables factorization, such that $C_1 ({\bf x})$ involves only $P_{1\perp}(\xi_1)$, $C_2 ({\bf x},{\bf y})$ involves only up to $P_{2\perp}(\xi_1,\xi_2)$, and so on. It is plausible that the Ansatz in Eq.~(\ref{eq:modA}) can be fine-tuned via a Bayesian analysis to have a good description of experimental data in hydrodynamic simulations. This would permit one to do so while keeping a simple relation with the nuclear structure.

\subsection{Diffractive photo-production of vector mesons}

\label{sec:dig1}

As originally pointed out in Ref.~\cite{Caldwell:2010zza}, the transverse two-body density, $P_{2\perp}(\xi_1,\xi_2)$, appears as well in the context of high-energy scattering, albeit in a different kind of process. This is the diffractive photo-production of vector mesons ($V$), where an incoming virtual photon ($\gamma^*$) interacts with a nuclear target via a virtual $q\bar q$ dipole, which is then produced to the final state as, e.g., a $\rho$ meson or a $J/\Psi$. 
In the small-$x$ formalism, the scattering amplitude for the production process is proportional to the Fourier transform of the dipole scattering amplitude  \cite{Kowalski:2006hc}
\begin{equation}
\label{eq:Aamp}
    \mathcal{A}^{\gamma^* A\rightarrow VA} \propto \int_{\bf b}  e^{-i{\bf b} \cdot {\bf \Delta} } N ({\bf r}, {\bf b}, x).
\end{equation}
Here ${\bf r}$ is the size of the scattering dipole, ${\bf b}$ is the distance between the dipole and the center of the nucleus, ${\bf \Delta}$ is the transferred transverse momentum, while $N ({\bf r}, {\bf b}, x)$ is the dipole scattering amplitude, usually taken in the IP-Sat formalism for a dipole scattering off a dense target \cite{Kowalski:2003hm},
\begin{equation}
\label{eq:dipamp}
     N ({\bf r}, {\bf b}, x) \propto 1 - e^{- {\bf r}^2 F({\bf r}, x) g ({\bf b})},
\end{equation}
where $F({\bf r}, x)\propto xg(x,\mu({\bf r}))$ carries the longitudinal momentum, $x$, and scale, $\mu({\bf r})$, dependence of the gluon distribution function, while $g({\bf b})$ is a phenomenological parametrization of the spatial density of gluons in the target. 

Consider now a nuclear target with a density of gluons given by the superposition of nucleon densities
\begin{equation}
    t({\bf b}) = \sum_{i=1}^{A} g({\bf b}-\xi_i).
\end{equation}
In the so-called weak field limit with $r^2 t({\bf b}) \ll 1$, Eq.~(\ref{eq:dipamp}) yields
\begin{equation}
    N ({\bf r}, {\bf b}, x) \propto t ({\bf b}),
\end{equation}
such that the scattering amplitude in Eq.~(\ref{eq:Aamp}) involves the Fourier transform of the nuclear configuration at the instant of scattering.

If the nuclear target breaks up or changes quantum state, the diffractive cross section has the form of a variance (\textit{incoherent} production, $t=-{\bf \Delta}^2$)
\begin{align}
\nonumber    & \frac{d\sigma^{\gamma^* A\rightarrow VA^*}}{d |t|} \propto \langle |\mathcal{A}|^2 \rangle  - | \langle \mathcal{A} \rangle  |^2 \propto \\
   & \int_{{\bf b}_1,{\bf b}_2} \left [  \langle t ({\bf b}_1) t ({\bf b}_2) \rangle  - \langle t({\bf b}_1) \rangle \langle t({\bf b}_2) \rangle  \right ] e^{-i{\bf \Delta\cdot({\bf b}_1-{\bf b}_2)}}.
\end{align}
The same convolutions of the previous sections appear (see also \cite{Blaizot:2022bgd}):
\begin{align}
    & \langle t({\bf b}_1) \rangle = A \int_{\xi_i} P_{1\perp}(\xi_i) g({\bf b}_1-\xi_i), \\
\nonumber    &  \langle t ({\bf b}_1) t ({\bf b_2}) \rangle = A \int_{\xi_i} P_{1\perp}(x_i) g({\bf b}_1-\xi_i)g({\bf b}_2-\xi_i) \\
    & \hspace{15pt} + (A^2-A) \int_{\xi_i\neq\xi_j} P_{2\perp} (\xi_i,\xi_j) g({\bf b}_1-\xi_i)g({\bf b}_2-\xi_j).
\end{align}
where the nucleon profile, $g({\bf b})$, is typically a 2D Gaussian with a width close to 0.4 fm ~\cite{Caldwell:2010zza,Mantysaari:2022ffw}.

Experimentally, these processes can be accessed either via electron-nucleus scattering or ultra-peripheral nucleus-nucleus scattering mediated by the Coulomb fields surrounding the colliding nuclei. In the context of ultra-peripheral collisions, a measurement of the \textit{coherent} cross section (involving $| \langle \mathcal{A} \rangle |^2 $, i.e., only the one-body density of the nucleus) for $\rho$ meson photo-production in ultra-peripheral $^{197}$Au+$^{197}$Au and $^{238}$U+$^{238}$U collisions has been recently achieved by the STAR collaboration \cite{STAR:2022wfe}. This has lead, in particular, to a precise determination of the neutron skin of $^{197}$Au. Even more recently, the first  measurement of the incoherent cross section for $J/\Psi$ photo-production in ultra-peripheral nucleus-nucleus collisions has been reported  by the ALICE collaboration \cite{ALICE:2023gcs}.
From the side of theory,   M\"antysaari \textit{et al.} \cite{Mantysaari:2023jny} have instead performed predictions for $J/\Psi$ photo-production using the deformed $^{238}$U nucleus as target. The resulting cross section as a function of the momentum transfer shows an enhancement in the low-$t$ region, corresponding to large spatial separations, that is consistent with the presence of a large quadrupole deformation. The same signal is observed as well with highly-deformed $^{20}$Ne targets.

It will be of fundamental importance, and a major challenge for nuclear physics in the future, to assess whether the same nuclear structure knowledge leads to a unified picture of different processes. In other words one should clarify whether the same nucleon density $P_{2\perp}(\xi_i, \xi_j)$ leads to a consistent understanding of the phenomenology of nuclei from low-energy experiments, to high-energy electron-nucleus collisions, to both ultra-central and ultra-peripheral nucleus-nucleus collisions.

\section{Numerical validation of the linearized approximation}

\label{sec:6}

Whether or not the present analysis has a phenomenological relevance depends on the validity of the linearized formulas, Eq.~(\ref{eq:epsnC2}) and Eq.~(\ref{eq:covC3}) in a realistic scenario of heavy-ion collisions. I perform now a check of their goodness in the IP-Jazma implementation. 

\subsection{Setup}

For this, a script in Python 3 has been developed to perform simulations of nuclear collisions. The script calculates on an event-by-event basis the energy density of the system starting from a simple model of the colliding ions. As in Fig.~\ref{fig:1}, I consider a mean-field description where the colliding nuclei are made of independent nucleons with an underlying particle density given by the Woods-Saxon profile:
\begin{equation}
\label{eq:WSlate}
    \rho(r, \theta, \Phi) \propto \frac{1}{1 + \exp \left (  \frac{r - R(\theta,\Phi)}{a} \right )}.
\end{equation}
To include the effect of many-body correlations, the half-width radius is expanded in (complex) spherical harmonics, $Y_l^m(\theta,\Phi)$, including the magnitude of the quadrupole deformation, $\beta_2$, the triaxiality parameter, $\gamma \in [0,60^\circ]$, and the magnitude of the octupole deformation, $\beta_3$, 
\begin{align}
     \nonumber &R(\theta,\Phi) = \\
     \nonumber  & R_0 \biggl \{ 1 + \beta_2 \biggl[ \cos \gamma ~Y_2^0(\theta,\Phi) + \sqrt{2} \sin \gamma~ {\rm Re}\bigl \{ Y_2^2(\theta,\Phi)\bigr \}  \biggr ] \\
     & ~~~~~ + \beta_3 Y_3^0(\theta,\Phi) \biggr\}.
\end{align}
A nucleus is then randomly oriented in space before the nucleons are sampled. The spherical one-body density results thus from an average over orientations:
\begin{equation}
    P_1(r_1,\theta_1,\Phi_1) = \frac{1}{8\pi^2} \int_{\Omega} \rho_\Omega(r_1,\theta_1,\Phi_1),
\end{equation}
where $\rho_\Omega(r,\theta,\Phi)$ denotes the intrinsic density rotated by a set of three Euler angles, $\Omega=(\alpha_1,\alpha_2,\alpha_3)$, in the lab frame. \footnote{For the average over orientations, we follow the standard convention where $\alpha_1$ is a rotation in the $(x,y)$ plane uniformly distributed between 0 and 2$\pi$, $\alpha_2$ is a rotation in the $(y,z)$ plane distributed such that $\cos(\alpha_2)$ is uniformly distributed between -1 and 1, $\alpha_3$ is a rotation in the $(x,y)$ plane uniformly distributed between 0 and 2$\pi$. This implies: 
\begin{equation}
\int_{\Omega} = \int_0^{2\pi} d\alpha_1 \int_{0}^{\pi} \sin \alpha_2 ~ d\alpha_2 \int_{0}^{2\pi} d\alpha_3  = 8 \pi^2 .  
\end{equation}
}
The two-body density of the system is instead obtained from the angular average of the two-point function of the intrinsic density, that is:
\begin{equation}
    P_2(r_1,\theta_1,\Phi_1,r_2,\theta_2,\Phi_2) = \frac{1}{8\pi^2} \int_{\Omega} \rho_\Omega(r_1,\theta_1,\Phi_1) \rho_\Omega(r_2,\theta_2,\Phi_2).
\end{equation}
Now, if the intrinsic density in Eq.~(\ref{eq:WSlate}) is spherical one has that:
\begin{equation}
    P_2(r_1,\theta_1,\Phi_1,r_2,\theta_2,\Phi_2) = P_1(r_1,\theta_1,\Phi_1,) P_1(r_2,\theta_2,\Phi_2),
\end{equation}
and analogously for the three-body density. On the other hand, correlations are produced as soon as deformation is included in the picture.

The simulations are performed on a transverse grid with $24\times24$ points, which ensures that the three-point correlations function (which has a total of $24^6 \approx 2\times10^8$ entries) can be easily stored on a laptop. I have tested that increasing the number of points has no visible influence on the computed quantities, which represent indeed global large-scale properties of the geometry of the sampled profiles. 
 Runs are performed with six different combinations of deformation parameters:
\begin{itemize}
    \item spherical nuclei: $\beta_2=0$, $\gamma=0$, $\beta_3=0$;
    \item prolate quadrupole-deformed nuclei: $\beta_2=0.5$, $\gamma=0$, $\beta_3=0$;
    \item octupole-deformed nuclei: $\beta_2=0$, $\gamma=0$, $\beta_3=0.5$;
    \item prolate quadrupole- and octupole-deformed nuclei: $\beta_2=0.5$, $\gamma=0$, $\beta_3=0.5$;
    \item Triaxial quadrupole- and octupole-deformed nuclei: $\beta_2=0.5$, $\gamma=30^\circ$, $\beta_3=0.5$;
    \item Oblate quadrupole- and octupole-deformed nuclei: $\beta_2=0.5$, $\gamma=60^\circ$, $\beta_3=0.5$.
\end{itemize}
For all these scenarios, I simulate: 
\begin{itemize}
    \item Collisions of nuclei with $A=192$, $R=6$ fm, $a=0.5$ fm. The grid size is $9.9\times9.9$ fm$^2$. The grid step is 0.825 fm. The results are shown in Fig.~\ref{fig:2}.
    
    \item Collisions of nuclei with $A=96$, $R=5$ fm, $a=0.5$ fm. The grid size is $9\times9$ fm$^2$. The grid step is 0.750 fm. The results are shown in Fig.~\ref{fig:3}.
    
    \item Collisions of nuclei with $A=48$, $R=4$ fm, $a=0.5$ fm. The grid size is $8.1\times8.1$ fm$^2$. The grid step is 0.675 fm. The results are shown in Fig.~\ref{fig:4}.
    
    \item Collisions of nuclei with $A=16$, $R=3$ fm, $a=0.5$ fm. The grid size is $7.2\times7.2$ fm$^2$. The grid step is 0.600 fm. The results are shown in Fig.~\ref{fig:5}.
\end{itemize}
For each setup, 50k collisions at zero impact parameter are simulated, leading to small enough statistical uncertainties.

After the sampling of coordinates, each nucleon is associated with a Gaussian transverse profile as in Eq.~(\ref{eq:Gaunucl}), which is evaluated up to a distance of $5w=2.5$ fm from its center. For both nuclei the transverse densities $t({\bf x})$ and $t^\prime({\bf x})$ are then constructed as the superposition of nucleon profiles. The energy density in one event is obtained as $\epsilon({\bf x})=t({\bf x})t^\prime ({\bf x})$. 

The observables analyzed in these calculations are those discussed in the previous sections (we consider hereafter $\langle E \rangle_{\rm ev}\equiv \langle E \rangle$):
\begin{itemize}
    \item ${\rm var}(E)$, ${\rm skew}(E)$, divided, respectively, by $\langle E \rangle^2$ or $\langle E \rangle^3$ to obtain dimensionless measures;
    \item $\varepsilon_n\{2\}$, for $n=2$ and $n=3$;
    \item ${\rm cov}(E,\varepsilon_n^2)$, for $n=2$ and $n=3$, divided by $\langle E \rangle$.
\end{itemize}
These quantities are evaluated either directly by calculating $E$ and $\mathcal{E}_n$ from the energy density field on an event-by-event basis (exact results), or perturbatively via Eq.~(\ref{eq:epsnC2}) and Eq.~(\ref{eq:covC3}) (perturbative results), for which the the connected $N$-point functions of the  density are computed. In the plots of Appendix~\ref{sec:appA}, the red symbols represent the exact evaluations, whereas the results displayed as black dashes are from the perturbative formulas. Further details are available in Appendix~\ref{sec:appA}.

\subsection{Results}

I first discuss the results related to collisions of nuclei with $A=192$, shown in Fig.~\ref{fig:2}. 

Concerning the fluctuations of $E$ (upper panels of Fig.~\ref{fig:2}), the perturbative result matches the Monte Carlo result as Eqs.~(\ref{eq:pt-var}) and (\ref{eq:pt-skew}) are exact. The expected nontrivial behavior is observed. Both ${\rm var}(E)$ and ${\rm skew}(E)$ are enhanced by $\beta_2$, though are not affected by an increase in the sole $\beta_3$, as also expected from Glauber-type calculations of the system size in the limit of central collisions \cite{Jia:2021qyu}. One sees in addition that variations of $\gamma$ have a subleading (though visible) impact on the integral of $C_2({\bf x},{\bf y})$ that determines ${\rm var}(E)$, while they yield a leading contribution to ${\rm skew}(E)$, determined by the connected three-point function, in qualitative agreement with the parametric expectation ${\rm skew} ( E ) = s_0 + s_1 \beta_2^3 \cos (3\gamma) $ \cite{Jia:2021qyu}, where $s_{0,1}$ are positive coefficients.

Moving on to $\varepsilon_n\{2\}$ (middle panels of Fig.~\ref{fig:2}), the linearized formula is essentially exact for collisions of spherical ions with $\beta_2=\beta_3=0$, which strongly motivates its use. The large increase of $\varepsilon_2\{2\}$ ($\varepsilon_3\{2\}$) due to $\beta_2=0.5$ ($\beta_3=0.5$), expected from the parametric relation $\varepsilon_n\{2\}^2 = c_0 + c_1 \beta_n^2$ \cite{Giacalone:2018apa,Giacalone:2021udy,Jia:2021tzt}, is precisely captured by the variation in the integral of $C_2({\bf x}, {\bf y})$. The linearized formula lies within 10\% of the exact result when $\varepsilon_n\{2\}$ is of order 0.3, in agreement with previous studies within independent-source models \cite{Gronqvist:2016hym}. I find in addition that the expected independence of $\varepsilon_n\{2\}$ on the value of $\gamma$ is verified by the estimate of Eq.~(\ref{eq:epsnC2}).

This demonstrates the impact of nuclear deformations on rms anisotropies in the context of the perturbative calculations. These results can lead to a better understanding of the implementation of nuclear structure in high-energy collisions. In the current modeling, for large nuclei one takes a deformed one-body density from a mean field calculation and uses it for the independent sampling of nucleon coordinates \cite{Ryssens:2023fkv}, as done in the present numerical study. If one could verify that the random rotation of an intrinsic shape leads to an appropriate description of the two-body density of the nuclear system, one could then conclude that the current use of the results of mean-field calculations in hydrodynamic simulations is justified. This is particularly relevant for octupole-deformed nuclei, such as $^{96}$Zr \cite{STAR:2021mii,Zhang:2021kxj}, whose deformation emerges from correlations on top of the mean field picture \cite{Rong:2022qez}. For such type of deformations, the literature suggests the following \cite{Bally:2021qys,Bally:2023dxi} in the theoretical framework of energy-density functional theory and the Projected Generator Coordinate Method (PGCM) \cite{Bender:2003jk}. After the symmetry restoration and the mixing of states, one can identify the most \textit{important} deformed point contributing to the correlated PGCM ground state. Then, one can use that information and determine a mean-field state in a Hartree-Fock-Bogoliubov calculation with deformations constrained to that point. The one-body density associated with the resulting state can subsequently be used as a randomly-oriented density of independent nucleons. To somehow validate this prescription, one possibility is thus to compute the one- and the two-body densities of the correlated PGCM wave function, inject into the formulas of this paper and see if the resulting $\varepsilon_n\{2\}$ matches that obtained from the randomly-oriented shape at the relevant deformed point. This would help motivate the current implementation of dynamical deformations in high-energy collisions.

I move to the lower panels of Fig.~\ref{fig:2}. For collisions of spherical nuclei, the exact results are
\begin{align}
\nonumber {\rm cov}(E,\varepsilon_2^2)/\langle E \rangle&=47(3)\times10^{-6}, \\
 {\rm cov}(E,\varepsilon_3^2)/\langle E \rangle&=38(4)\times10^{-6}   ,
\end{align}
while the perturbative expressions lead to
\begin{align}
    {\rm cov}(E,\varepsilon_2^2)/\langle E \rangle & =44\times10^{-6}, \\
    \nonumber  {\rm cov}(E,\varepsilon_3^2)/\langle E \rangle & = 15\times10^{-6}.
\end{align}
The latter value points, thus, to a shortcoming of the perturbative formula. The parametric expectation for the covariance of $E$ and $\varepsilon_2^2$ is ${\rm cov} ( E, \varepsilon_2^2 ) = s^\prime_0 - s^\prime_1 \beta_2^3 \cos (3\gamma) $ \cite{Jia:2021qyu}. An analogous formula is likely to hold as well for ${\rm cov} ( E, \varepsilon_3^2 )$. The Monte Carlo results predict indeed a strong suppression of the covariance due to increasing $\beta_2$ \cite{Giacalone:2019pca,Giacalone:2020awm,Jia:2021wbq}. This is captured by the linearized formula, showing that this change involves indeed the connected three-point function of the density field. Second, one observes the leading contribution of $\gamma$ to this observable, with the correlator essentially flipping sign as one moves from prolate nuclei with $\gamma=0$ to oblate nuclei with $\gamma=60^\circ$. This effect is again captured by the connected three-point function, which does not however lead to a quantitative description of the exact ${\rm cov} (E,\varepsilon_2^2)$ result. One notes in addition that the suppression of ${\rm cov}(E,\varepsilon_3^2)$ is observed only when both $\beta_2$ and $\beta_3$ are turned on. A residual dependence on $\gamma$ of ${\rm cov}(E,\varepsilon_3^2)$ is also partially captured by the perturbative formula.

Figures \ref{fig:3} to \ref{fig:5} show results for collisions of nuclei with lower mass numbers. They illustrate the breakdown of the linearized expression as soon as the fluctuations of the system are dominated by the small nucleon number. In Fig.~\ref{fig:5}, ${\rm var}(E)$ and ${\rm skew} (E)$ for $A=16$ have little residual dependence on deformation parameters. The effect of $\gamma$ is in particular largely washed out. Concerning $\varepsilon_n\{2\}$, the perturbative formulas provide a good description of the Monte Carlo data down to $A=48$. At $A=16$, the value of the rms eccentricities is above 0.3 already for spherical nuclei, which engenders a significant error. Here the effect of the deformations is also largely washed out by the small nucleon number. However, small effects are captured by the perturbative calculations, in particular:
\begin{equation}
    \frac{\varepsilon_n\{2\}_{\beta_n=0.5}}{\varepsilon_n\{2\}_{\beta_n=0}}\biggl|_{\rm exact} \approx \frac{\varepsilon_n\{2\}_{\beta_n=0.5}}{\varepsilon_n\{2\}_{\beta_n=0}}\biggl|_{\rm perturbative},
\end{equation}
meaning that these ratios cancel much of the theoretical error induced by the linearization. This may be relevant in the study of the aforementioned $^{20}$Ne nucleus. The peculiar shape of $^{20}$Ne leads to a $\approx 10\%$ enhancement of $v_2\{2\}$ in $^{20}$Ne+$^{20}$Ne collisions relative to $^{16}$O+$^{16}$O collisions \cite{inprep}. This comes from the  ratio of the initial $\varepsilon_2\{2\}$ taken between these two systems, which could be thus captured by the perturbative formula from the computation of the two-body densities, $P_{2\perp}(\xi_i,\xi_j)$, which should be affordable to any \textit{ab initio} framework of nuclear structure.

Finally, the observables ${\rm cov}(E,\varepsilon_n^2)$ are more strongly impacted by the lowering of the nucleon number. For $A=16$, the mild dependence on the deformation parameters shown by the exact result is entirely lost in the perturbative formulas.  These results may be improved in future by adding an extra power of $\delta \epsilon ({\bf x})$ in the perturbative expansion.  Alternatively, one could think about other expansion schemes which may be more suited to address small systems \cite{Floerchinger:2020tjp,Borghini:2022iym}.

\section{Conclusion \& Outlook}

\label{sec:7}

I have presented a field-theoretical approach to energy-density correlations in the QGP induced by many-body correlations of nucleons in the wave functions of the colliding nuclei. The energy deposition formula of Eq.~(\ref{eq:jazma}) provides a simple and yet realistic description of the energy density at $\tau=0^+$. If the density of gluons in a nucleus at high energy is a superposition of nucleonic profiles, one obtains straightforward relations between $N$-point functions of the energy density field and $N$-nucleon density distributions in the scattering nuclei.  Combined with the good quality of the linearized approach to energy-density fluctuations, especially for collisions of $A>48$ nuclei, this demonstrates that multi-nucleon correlations in the initial states and multi-hadron correlations in the final states are closely connected.  

This provides a formal justification for the impact of features such as nuclear deformations on the outcome of nuclear collisions at high energy. Hopefully, it will serve as a starting point towards the systematic implementation of  different  nuclear interactions on model calculations of such processes. In the Monte Carlo study of Sec.~\ref{sec:4},   a simple model of independent nucleons with a deformed intrinsic density is employed. However, tabulated one-, two- and three-nucleon densities, following Eqs.~(\ref{eq:P2t}) and (\ref{eq:P3t}), if computed systematically in low-energy theory, could be directly employed in the equations presented in this paper. High-energy observables may reveal different sensitivities to the parameters of the nuclear interaction compared to the observables studied in low-energy experiments. This would in turn demonstrate low- and high-energy nuclear experiments as complementary means to advance our knowledge of the strong nuclear force. 

In addition, in the present formalism high-energy physics and low-energy nuclear structure are essentially decoupled. In practice, though, the nuclear two-body density, $P_{2\perp}(\xi_i,\xi_j)$, might be modified by the strong Lorentz boost on length scales comparable to the nucleon size, $w\approx0.5$ fm. We have no knowledge at present regarding how such modifications may look like. Precise measurements of the incoherent cross section discussed in Sec.~\ref{sec:dig1} performed at $t$ scales intermediate between 1 GeV$^2$ and the pion mass squared ($m_\pi^2 \approx 0.02$ GeV$^2$) represent a promising avenue to shed light on these unexplored properties of nuclei at high energy. They may be achieved from future high-statistics $^{208}$Pb+$^{208}$Pb runs at the CERN LHC as well as at the EIC.

Finally, this article discusses symmetric collisions of even-even nuclei at zero impact parameter. Generalization to different situations should be straightforward, and will be the subject of follow-up works. Furthermore, several multi-particle correlations measured in heavy-ion collision probe the non-Gaussianity of fluctuations through the connected four-point function of the energy density field \cite{Bhalerao:2019fzp},  whose analysis is left for a future study.

\section{Acknowledgements}
I acknowledge discussions with the participants of the INT Program INT-23-1a, ``Intersection of nuclear structure and high‐energy nuclear collisions'', and the hospitality of the Institute for Nuclear Theory, Seattle, where this work was initiated. 
I thank Vittorio Som\`a for useful comments on the manuscript.
I also thank Benjamin Bally, Daniel Brandenburg, Abhay Deshpande, Thomas Duguet, Jean-Paul Ebran, Jiangyong Jia, Sebastian K\"onig, Tyler Kutz, Dean Lee, Elena Litvinova, Matthew Luzum, Hadi Mehrabpour, Jean-Yves Ollitrault, Eli Piasetzki, Govert Nijs, Jennifer Rittenhouse West, Wouter Ryssens, Bj\"orn Schenke, Vittorio Som\`a, Anna Stasto, Kong Tu, Wilke van der Schee, Chunjian Zhang, Wenbin Zhao for fruitful discussions.
This research is funded by the DFG (German Research Foundation) -- Project-ID 273811115 -- SFB 1225 ISOQUANT.

\begin{appendices}

\section{Numerical results and figures}

\label{sec:appA}

\begin{figure*}[h!]
    \centering
    \includegraphics[width=.9\linewidth]{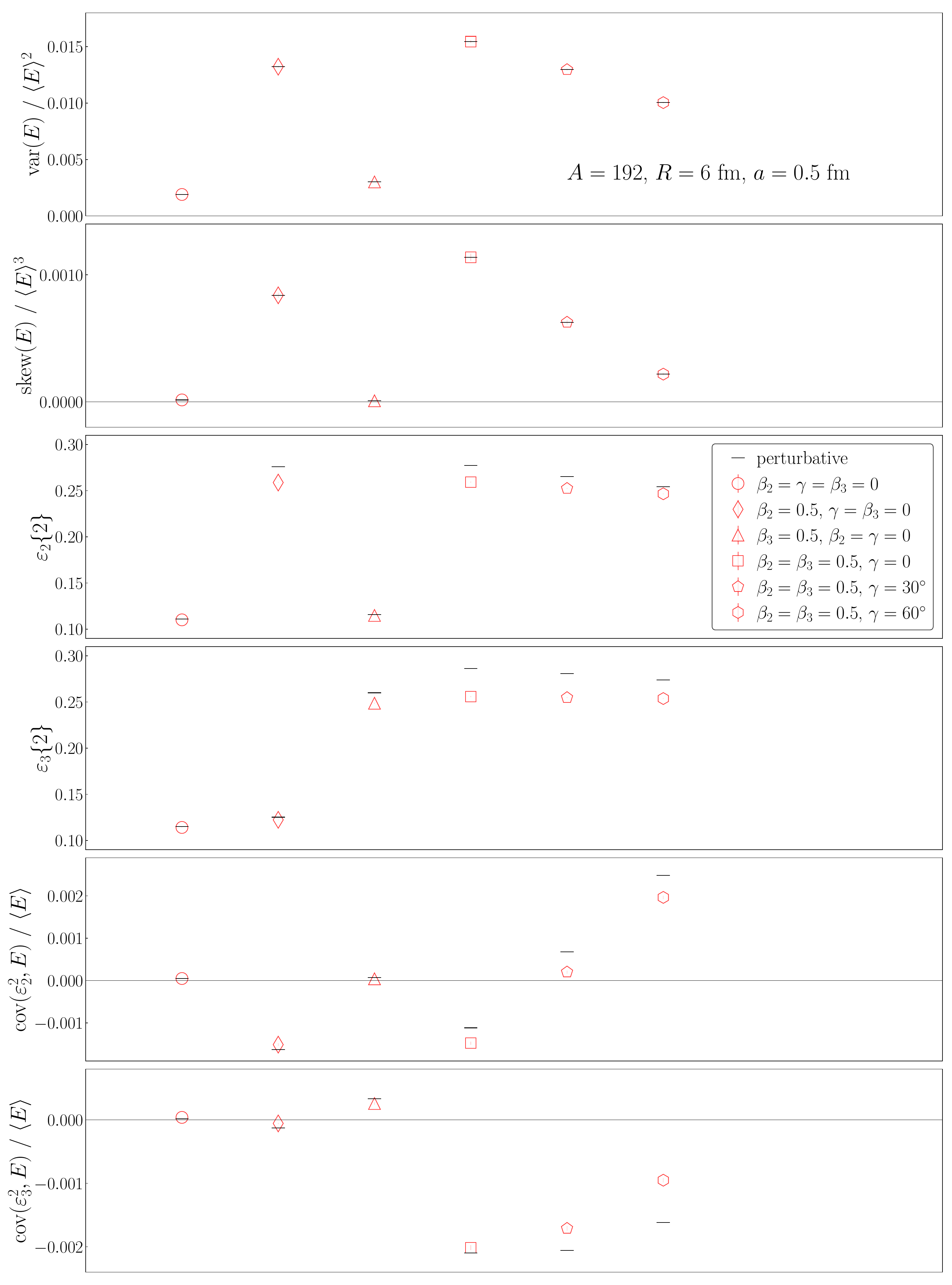}
    \caption{Results for nuclei with $A=192$, $R=6$ fm, $a=0.5$ fm. Statistical error bars are smaller than the size of the symbols.}
    \label{fig:2}
\end{figure*}

\begin{figure*}[h!]
    \centering
    \includegraphics[width=.9\linewidth]{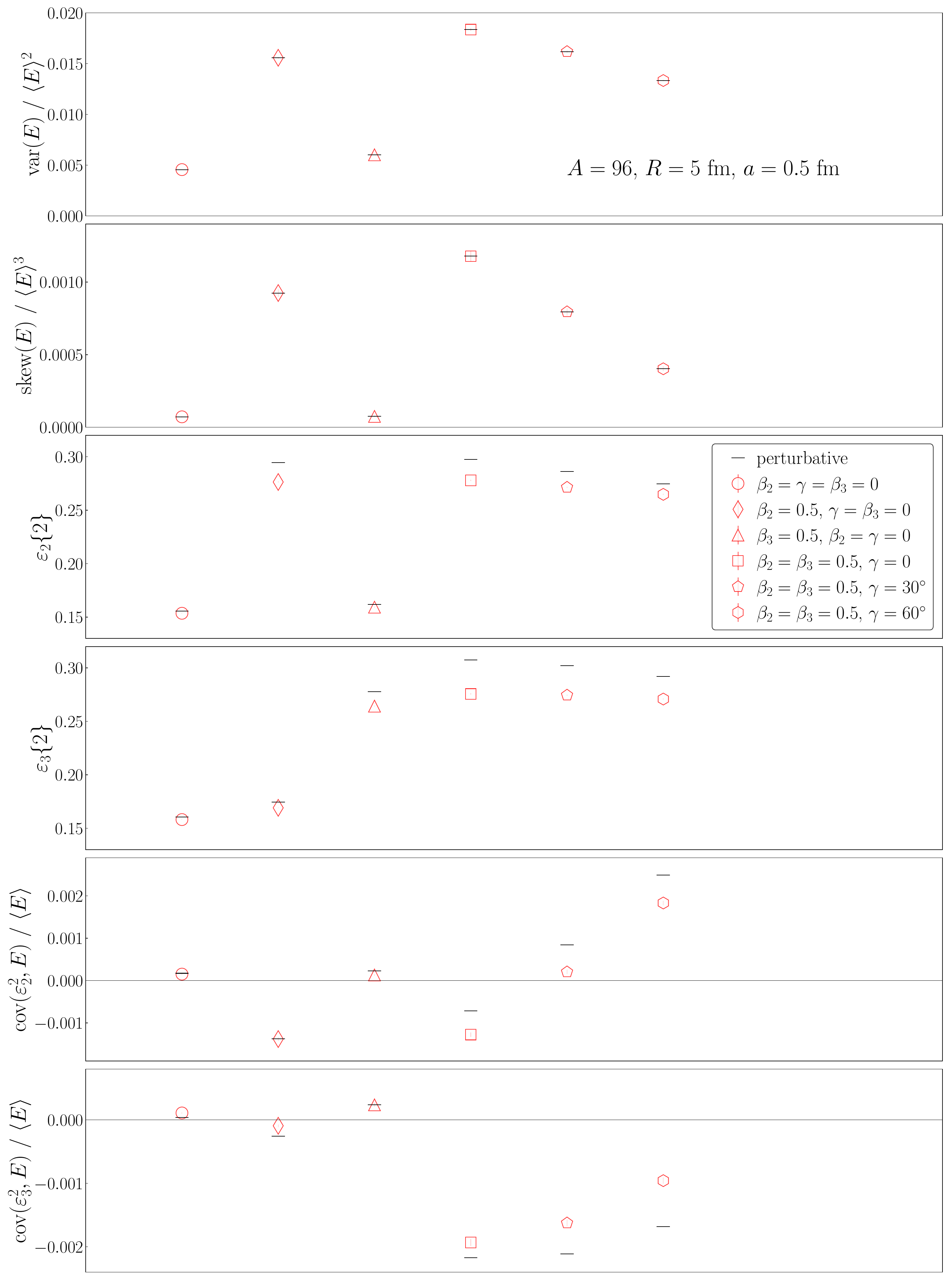}
    \caption{Same as Fig.~\ref{fig:2}, but for collisions of nuclei with $A=96$, $R=5$ fm, $a=0.5$ fm.}
    \label{fig:3}
\end{figure*}

\begin{figure*}[h!]
    \centering
    \includegraphics[width=.9\linewidth]{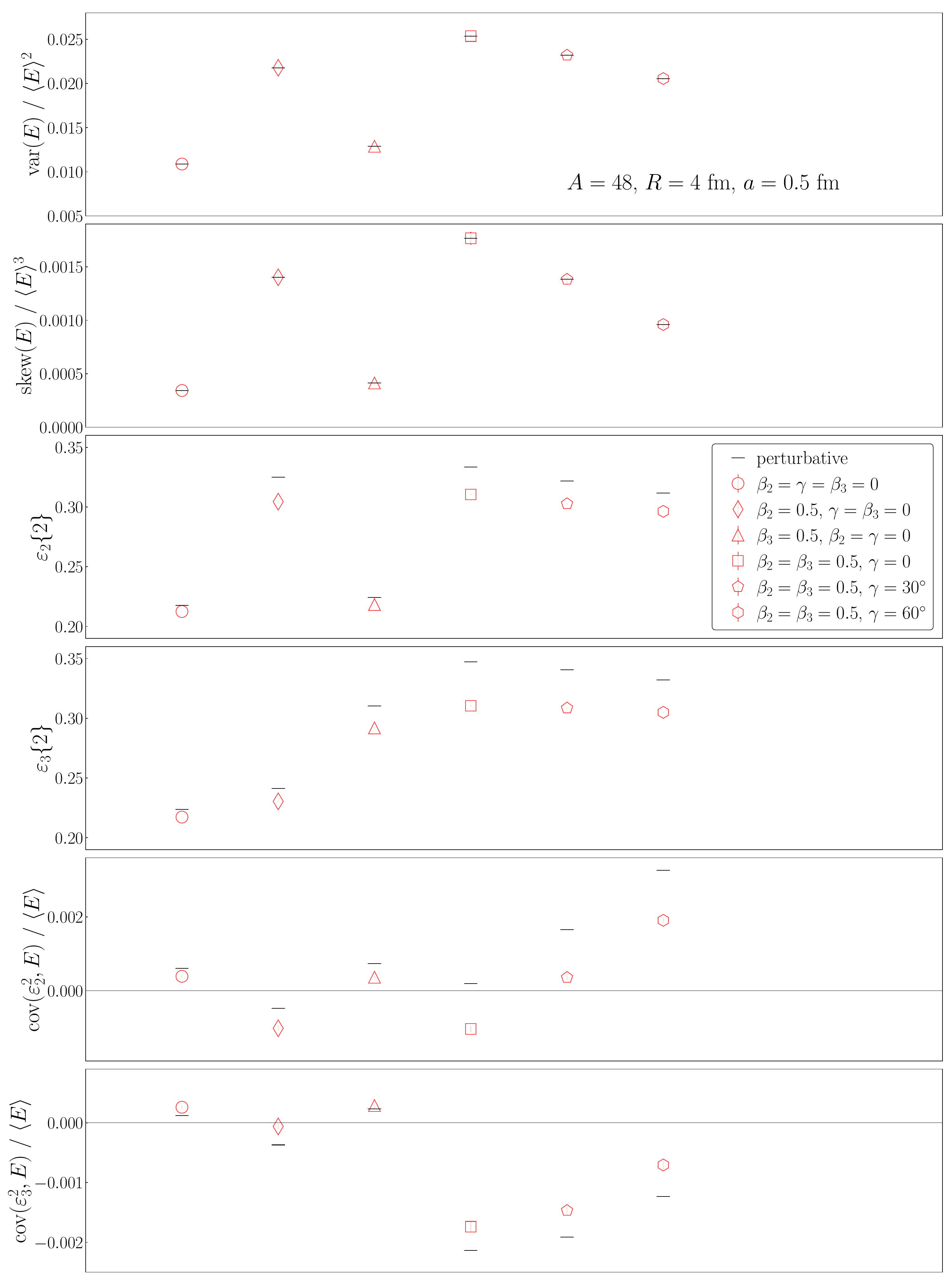}
    \caption{Same as Fig.~\ref{fig:2}, but for collisions of nuclei with $A=48$, $R=4$ fm, $a=0.5$ fm.}
    \label{fig:4}
\end{figure*}

\begin{figure*}[h!]
    \centering
    \includegraphics[width=.9\linewidth]{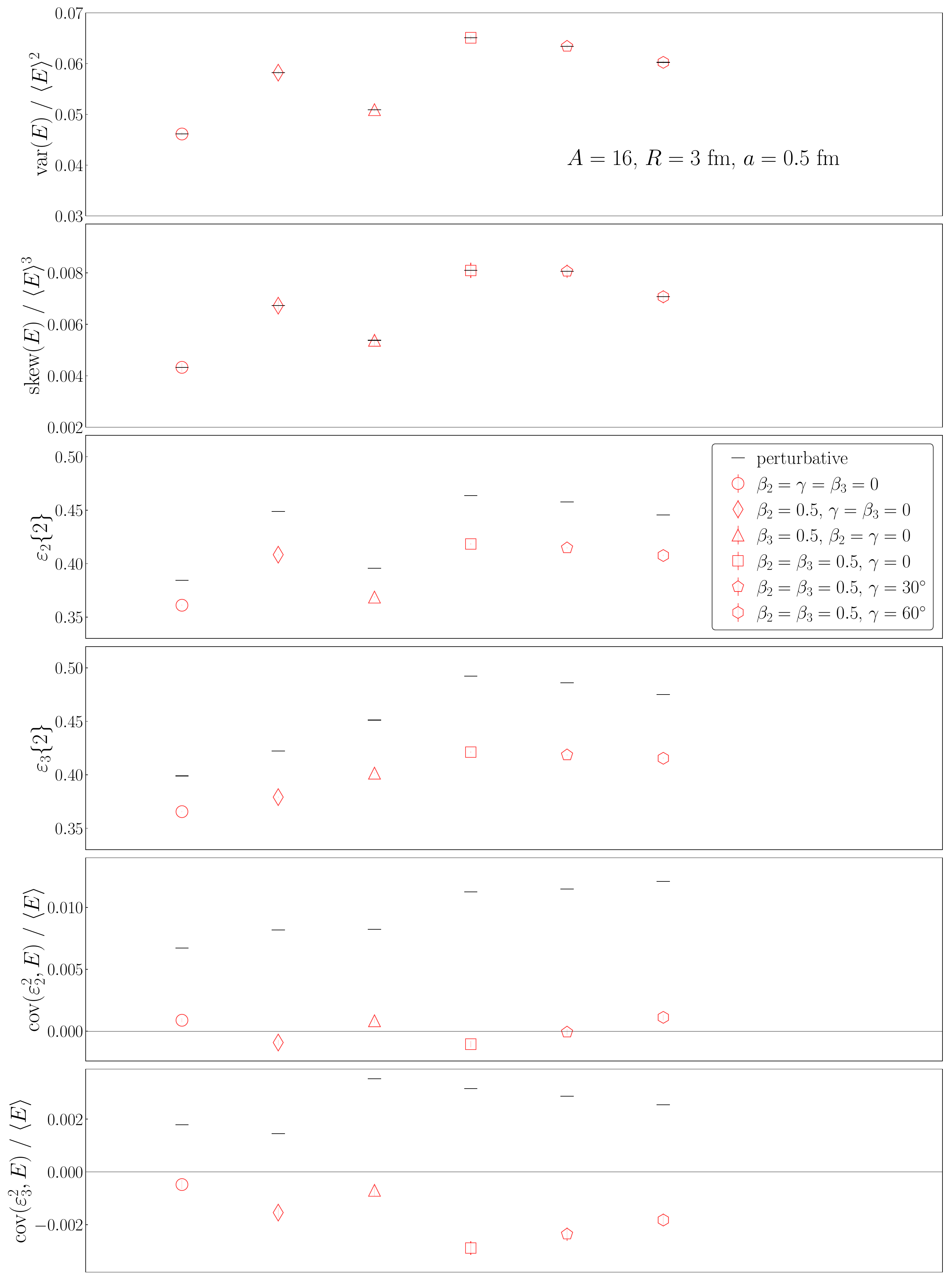}
    \caption{Same as Fig.~\ref{fig:2}, but for collisions of nuclei with $A=16$, $R=3$ fm, $a=0.5$ fm.}
    \label{fig:5}
\end{figure*}

 I show plots with the results of the numerical simulations discussed in Sec.~\ref{sec:4}. The figures contain results for collisions of nuclei presenting, respectively, $A=192$ (Fig.~\ref{fig:2}), 96 (Fig.~\ref{fig:3}), 48 (Fig.~\ref{fig:4}), 16 (Fig.~\ref{fig:5}), and different nuclear geometry parameters. Each figure has 6 panels, corresponding to the total number of analyzed observables. The results displayed as symbols correspond to exact evaluations obtained from the Monte Carlo simulations. For each observable, the calculations have been performed for 6 different choices of nuclear deformation parameters, which correspond to different marker styles.   The results displayed as horizontal lines are instead obtained from the perturbative calculations involving the correlations functions of the energy density field. I refer to Sec.~\ref{sec:4} in the main text for the discussion and the interpretation of the numerical results.

\end{appendices}

\newpage 

\newpage 

\newpage 

\newpage
~

\newpage 

~
\newpage 
~

\newpage
~

\newpage
~

\newpage
~
\newpage

\newpage 
~

\newpage
~

\end{document}